\documentclass{article}
\usepackage{frExamplee}
\usepackage{amsmath}
\usepackage{graphicx}
\usepackage{subcaption}
\usepackage{xcolor}
\usepackage{soul}

\begin{document}

\title{\vspace*{0.25in} A hierarchical adaptive nonlinear model predictive control approach for maximizing tire force usage in autonomous vehicles\\
\thanks{J. Dallas, M. Thompson, J. Goh, and A. Balachandran are with  the Extreme Vehicle Dynamics Control group, Toyota Research Institute, Los Altos, CA, 94022.\\ $^*$Corresponding author: james.dallas@tri.global\\
$^\dagger$ Authors contributed equally%
        }
}

\author{James~Dallas$^{\dagger}$,
        Michael~Thompson$^{\dagger}$,
        Jonathan~Y.M.~Goh$^*$,
        Avinash~Balachandran
}

\maketitle

\begin{abstract}
The ability to reliably maximize tire force usage would improve the safety of autonomous vehicles, especially in challenging edge cases. However, vehicle control near the limits of handling has many challenges, including robustly contending with tire force saturation, balancing model fidelity and computational efficiency, and coordinating inputs with the lower level chassis control system. This work studies Nonlinear Model Predictive Control for limit handling, specifically adapting to changing tire-road conditions and maximally allocating tire force utilization. We present a novel hierarchical framework that combines a single-track model with longitudinal weight transfer dynamics in the predictive control layer, with lateral brake distribution occurring at the chassis control layer. This vehicle model is simultaneously used in an Unscented Kalman Filter for online friction estimation. Comparative experiments on a full-scale vehicle operating on a race track at up to 95\% of maximum tire force usage demonstrate the overall practical effectiveness of this approach.  

\end{abstract}

\section{Introduction}
The ability to fully use the force generation capabilities of a vehicle can greatly improve the safety of autonomous vehicles. For example, in \cite{wurts2020,DALLASIFAC,WurtsACC}, a collision imminent steering algorithm was developed to perform an evasive lane change when collision could not be avoided by braking alone. However, doing so required the vehicle to operate near tire force saturation, where the closed-loop behavior of the vehicle is strongly influenced by the level of model fidelity used \cite{LiuModel}. The objective of minimizing laptime in racing scenarios provides an opportunity to further explore the role of model fidelity in safely and reliably controlling autonomous vehicles at the limits. For an autonomous controller to extract full performance out of a vehicle, the vehicle model must capture complex vehicle dynamics, and the controller must have precise knowledge of the current environment. This work addresses both of these needs.

Model Predictive Control (MPC) has drawn interest in limit handling applications as the vehicle dynamics, constraints, and costs can be intuitively encoded in a receding horizon manner \cite{BROWN2017307, Schwarting}. In doing so, MPC has the capability to replan trajectories online to balance objectives and constraints to account for modeling error and changing environment conditions. While the state-of-the-art has explored MPC for high level planning and control, MPC often ignores important aspects of the underlying chassis controller such as allocating brake balance. This raises an important question: could performance and safety be improved by subsuming some of the chassis control functionality into the higher level MPC model?

Bringing elements of chassis control functionality into the high-level MPC accentuates an important trade-off -- the vehicle model must sufficiently capture the complex dynamics that occur at the handling limits while balancing computational complexity for real-time operation. Various MPC formulations have been developed in an effort to balance model complexity and efficiency. Approaches have varied in model fidelity from linear MPC which can reduce computational effort at the expense of modeling error \cite{Katriniok,Turri}, to Nonlinear MPC (NMPC) accounting for road topology, nonlinear dynamics, and force constraints in racing applications \cite{LaurenseThesis}. Furthermore, various approaches addressing computational effort of NMPC have focused on extending horizons through cascaded approaches \cite{LaurenseThesis,LaurenseACC,Laurense2018AVEC}, pseudospectral methods \cite{FebboThesis}, and parallelization \cite{wurts2020}. While these approaches address the balance of model fidelity and computational complexity, efficiently bringing low level chassis control into the higher level optimizer -- and demonstrating its practical utility -- remains an open question.

Even with a high fidelity model, knowledge of the evolving tire-road interaction is needed to extract the full potential of an autonomous controller. For example, due to the sensitivity of tire forces to friction in limit handling scenarios, even just a deviation of 2\% can lead to failure \cite{LaurenseACC}, demonstrating the importance of accurately modeling the tire-road interaction.  Various approaches to address tire parameter adaptation have been proposed, including adaptive linear and nonlinear tracking controllers \cite{Chen-BC, Borrelli2005, Falcone}, and adaptive coupled trajectory planning and tracking formulations \cite{DallasIEEE, DALLASIFAC, WurtsACC}.  The latter of these examples was demonstrated in simulation and decreased the sensitivity to changing tire parameters by allowing for online planning that can account for updated parameters that are unknown \textit{a priori}, preventing infeasibility of offline plans. However, extensive experimental validation of a UKF based friction estimator, as well as including the impact of topology and force coupling in a high-fidelity estimation model, has yet to be addressed \cite{DALLAS202011}.

This work builds upon the state-of-the-art with a hierarchical adaptive NMPC approach that subsumes longitudinal brake balance into the predictive control layer, but delegates lateral brake balance to the chassis layer. This permits using the single-track assumption in the optimization problem, enabling reduced complexity and longer horizons for stability, whilst still maximizing tire force usage on all four wheels during limit braking scenarios. To extract the full potential of this approach, it is combined with a novel application of a UKF based friction estimation algorithm which extends that of \cite{DALLAS202011}. 

Specifically, first-order longitudinal load transfer dynamics and steady-state lateral weight transfer are modeled to account for the evolving force potential at each tire. This enables optimal allocation of brake torque at each axle. A low-level routine apportions lateral brake distribution based on feedforward lateral acceleration from the reference trajectory, and in turn, the NMPC optimization problem accounts for the induced yaw effect from braking. Next, an Unscented Kalman filter (UKF) is utilized to estimate tire-road friction in real time.  Lastly, the NMPC prediction model is updated with the estimated coefficient of friction to adapt to uncertainties in real time.

Experimental validation on a race track at an imposed limit of $95\%$ of the available friction, with and without online UKF estimation, shows the efficacy of this approach. Furthermore, comparative experiments between dynamic brake balance and static distributions showcase the importance of including this capability in the predictive control layer.

The vehicle model is presented in Section II. The NMPC formulation is presented in Section III, and the UKF friction estimator is discussed in Section IV.  Experimental setup is given in Section V and insights drawn from the NMPC formulation are discussed in Section VI. Finally, Sec. VII draws conclusions and discusses future work.

\section{Vehicle Model}
\subsection{Bicycle Model} \label{sec:bicycle}
\begin{figure}[bp]
    \centering
    \includegraphics[width=0.44\textwidth]{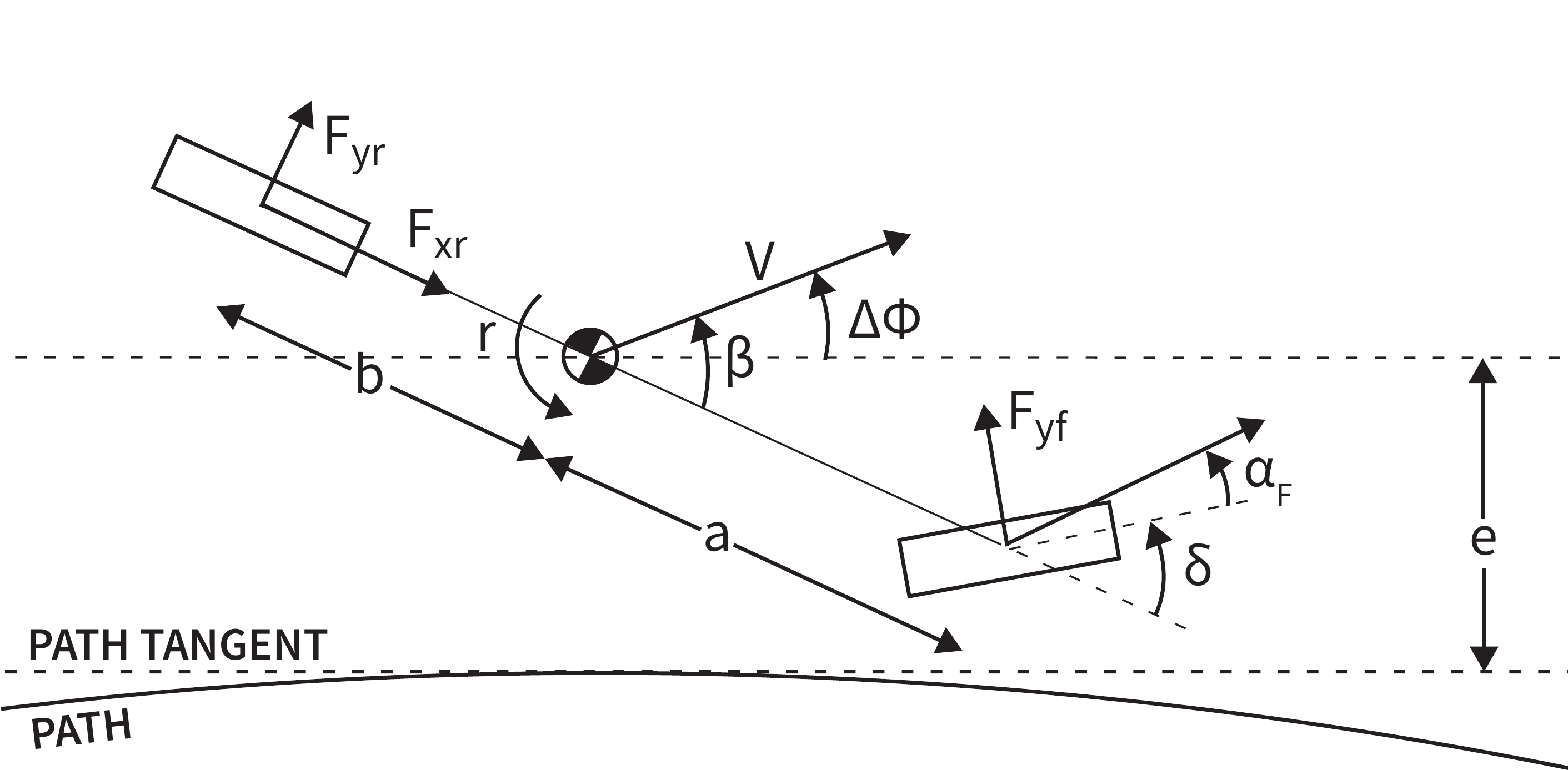}
    \caption{Single track bicycle model.}
    \label{fig:bike_model}
\end{figure}

The vehicle model used for the NMPC controller and for generating reference trajectories is given by the single-track dynamic bicycle model in a curvilinear coordinate system \cite{Goh2019TowardAV,GohThesis}, illustrated in Fig. \ref{fig:bike_model}. There are a total of 11 vehicle states in this model, including a state representing the transient load shift occurring from the first order longitudinal weight transfer model described in Sec. \ref{sec:weight_long}. The states are described in Eq. \ref{eqn:symbols}. 

\begin{equation}
    x = 
    \begin{bmatrix}
    r \\
    V \\
    \beta \\
    \omega_r \\
    e \\
    \Delta\phi \\
    dF_z \\
    \delta \\
    \tau \\
    \tau_{brake,f} \\
    \tau_{brake,r} 
    \end{bmatrix}
    =
    \begin{bmatrix}
    \textit{Yaw rate} \\
    \textit{Velocity} \\
    \textit{Sideslip} \\
    \textit{Rear wheelspeed} \\
    \textit{Lateral error} \\
    \textit{Course error} \\
    \textit{Longitudinal weight transfer state} \\

    \textit{Roadwheel angle} \\
    \textit{Engine torque} \\
    \textit{Front brake torque} \\
    \textit{Rear brake torque} \\
    \end{bmatrix}
    \label{eqn:symbols}
\end{equation}

To easily encode actuator slew rate constraints, the 4 inputs to the model are the rate of change of roadwheel angle (steering), engine torque, and the front and rear brake torques:
\begin{equation}
    u= \begin{bmatrix}
    \dot{\delta}\\
    \dot{\tau} \\
    \dot{\tau}_{brake,f} \\
    \dot{\tau}_{brake,r}
    \end{bmatrix}
    =
    \begin{bmatrix}
    \textit{Roadwheel angle rate} \\
    \textit{Engine torque rate} \\
    \textit{Front brake torque rate} \\
    \textit{Rear brake torque rate}
    \end{bmatrix}
\end{equation}

The state derivatives describing this vehicle model are given as:

\begin{equation} \label{eq:bicycle}
    \dot{x} = \begin{bmatrix}
    \frac{aF_{yf}cos(\delta) +aF_{xf}sin(\delta)-bF_{yr} + \tau_{bb}}{I_z} \\
    
    \frac{(-F_{yf}sin(\delta - \beta)+F_{xf}cos(\delta - \beta)+(F_{yr}+F_{gy})sin(\beta)+(F_{xr}+F_{gx})cos(\beta))}{m} \\
    
    \frac{(F_{yf}cos(\delta - \beta)+F_{xf}sin(\delta - \beta)+(F_{yr}+F_{gy})cos(\beta)-(F_{xr}-F_{gx})sin(\beta))}{mV}-r \\
    
    \frac{r_w(\tau_w -F_{xr}r_w)}{I_w} \\

    Vsin(\Delta\phi) \\
    
    \dot{\phi} - \kappa_{ref}\frac{Vcos(\Delta\phi)}{1-\kappa_{ref}e} \\
    
    -k \left(dF_z - \frac{h_{cg}}{a+b} F_{xnet}\right) \\

    \dot{\delta} \\

    \dot{\tau} \\
    
    \dot{\tau}_{brake,f} \\
    \dot{\tau}_{brake,r}\\
    
    \end{bmatrix}
\end{equation}

Where $a$ and $b$ are the distance from the center of gravity to the front and rear axles, respectively, $h_{cg}$ is the center of gravity height, $r_w$ is the tire radius, $m$ is the vehicle mass, and $I_z$ and $I_w$ are the yaw moments of inertia for the vehicle and lumped rear axle, respectively. The longitudinal and lateral tire forces are given as $F_{xf,r}$ and $F_{yf,r}$, for the front and rear tires respectively, $\tau_w$ is the torque at the wheel, $\tau_{bb}$ is the moment created from the lateral brake balance discussed in Sec. \ref{sec: lateral_bb}, $F_{gx}$ and $F_{gy}$ are the gravitational forces in the longitudinal and lateral directions due to road topology described in Sec. \ref{sec:topology}, and the longitudinal weight transfer model is further described in Sec. \ref{sec:weight_long}.  $\kappa_{ref}$ is the reference curvature and

\begin{equation}
    \dot{\phi} = \dot{\beta} + \dot{r}
\end{equation}

is the rate of rotation of the vehicle's velocity vector.

\subsection{Tire Model} \label{sec:tire}
The forces $F_{xf,r}$ and $F_{yf,r}$ are modeled by an isotropic coupled slip Fiala brush tire model, similar to that described in \cite{svendenius2007tire}.  This is given as

\begin{equation}
\begin{bmatrix}
F_y \\
F_x
\end{bmatrix}
= F_{total}
\begin{bmatrix}
\frac{-tan(\alpha)}{\sigma} \\
\frac{\kappa}{\sigma}
\end{bmatrix}
\label{eq:forces}
\end{equation}

where $\kappa$ is the slip ratio, $\alpha$ is the slip angle, and $\sigma$ is the combined slip given as

\begin{equation}
    \sigma = \sqrt{tan(\alpha)^2 + \kappa^2}
\end{equation}

and $F_{total}$ is given as

\begin{align*}
F_{total} &=  
\begin{cases}
C_f\sigma - \frac{C_f^2 \sigma^2}{3\mu F_z}  + \frac{C_f^3\sigma^3}{27(\mu F_z)^2} & |\sigma| < \sigma_{sl}\\
\mu F_z & |\sigma| > \sigma_{sl}\\
\end{cases}
\end{align*}

Where $C_f$ is the cornering stiffness,  $\mu$ is the coefficient of friction, $F_z$ is the normal load, and $\sigma_{sl}$ is the maximum combined slip where saturation occurs:

\begin{equation}
    \sigma_{sl} = arctan(3\mu F_z/C_f)
\end{equation}

\subsection{Road Topology} \label{sec:topology}
The effects of road topology are incorporated as in \cite{LaurenseThesis, SubositsThesis}.  The effect of topology on the normal load at the front and rear axle is given as:

\begin{equation}
    F_{zf, topology} = \frac{b}{a+b}m(g\cos(\theta)\cos(\psi) + A_v)
\end{equation}

\begin{equation}
    F_{zr, topology} = \frac{a}{a+b}m(g\cos(\theta)\cos(\psi) + A_v)
\end{equation}

where $\theta$ and $\psi$ are the road grade and bank, respectively. The centripetal acceleration due to vertical curvature, $A_v$, is given as:

\begin{equation}
    A_v = \left(-\frac{d\theta}{ds}cos(\psi)-\kappa sin(\psi)cos(\theta)\right)(\dot s)^2
\end{equation}

Road grade and bank also contribute components of gravitational acceleration along the vehicle's longitudinal and lateral direction.  This is given respectively as:
\begin{equation}
    F_{gy} = -mg\cos(\theta)sin(\psi)
\end{equation}

\begin{equation}
    F_{gx} = mg\sin(\theta)
\end{equation}

\subsection{Load Transfer} \label{sec:weight_long}
As seen, in Sec. \ref{sec:tire}, the tire force -- and importantly, its maximum -- depends on the normal force at each tire. This makes accurate, but computationally efficient, modelling of load transfer dynamics crucial to on-road performance. Inspired by experimental data, we use a simple but accurate first order model for the longitudinal load transfer dynamics.  This is given as:

\begin{equation} \label{eqn:wt_model}
    \dot{dF_z} = -k \left(dF_z - \frac{h_{cg}}{a+b} F_{xnet}\right)
\end{equation}

Where $k$ is a constant, $dF_z$ is the load transferred from the front to rear axle due to the acceleration of the car, and the net longitudinal force, $F_{xnet}$, is given as 

\begin{equation}
    F_{xnet} = F_{xr} + F_{xf}cos(\delta) - F_{yf} sin(\delta) + F_{gx}
\end{equation}

Hence, the load on the front and rear axles are given respectively as:

\begin{equation}
    F_{zf} = F_{zf, topology} - dF_z
\end{equation}

\begin{equation}
    F_{zr} = F_{zr, topology} + dF_z
\end{equation}

\begin{figure}
    \centering
    \includegraphics[scale = 0.75]{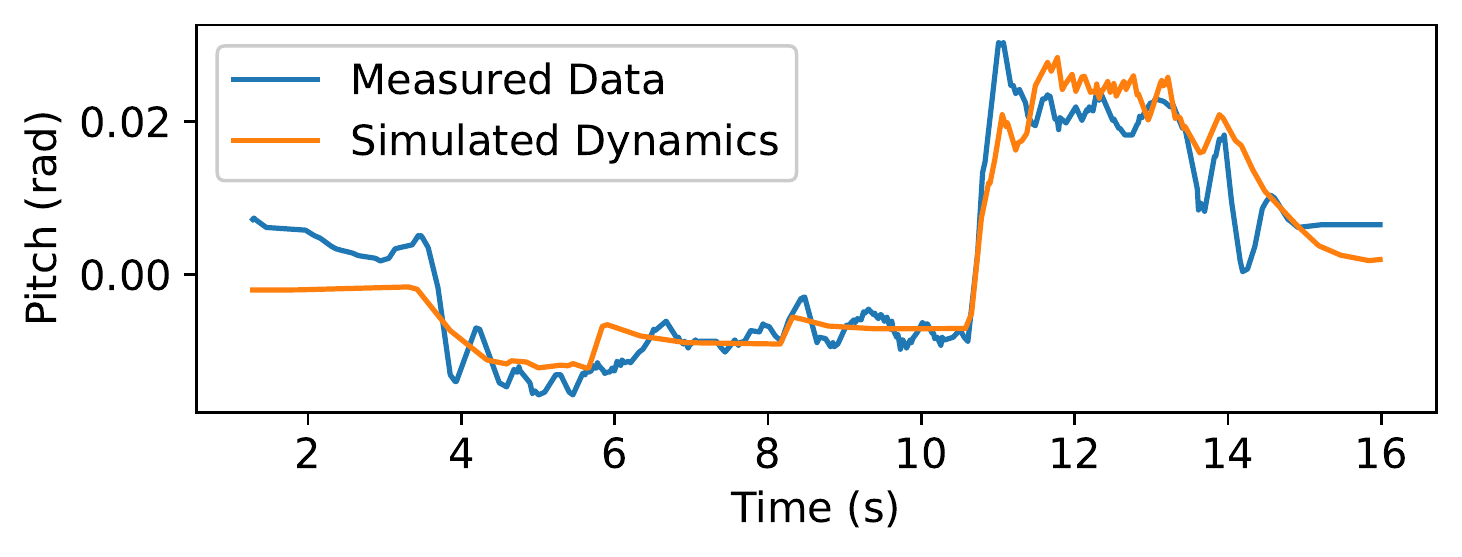}
    \caption{Pitch response of test vehicle during acceleration (4s to 10s) and braking (11s to 14s). This data is used to calibrate the weight transfer model, specifically $k$ in Eq. \eqref{eqn:wt_model}.}
    \label{fig:wt_calibration}
\end{figure}

The weight transfer model time constant $k$ is calibrated based on data taken from the test vehicle (Sec. \ref{sec: exp_veh}). First, the effective pitch stiffness was calculated from measurements of pitch angle during extended periods of constant acceleration and constant braking. Transient pitch behavior, measured during step changes in acceleration and braking, is then used to estimate $k$ from Eq. \eqref{eqn:wt_model}. Fig. \ref{fig:wt_calibration} shows an example of measured pitch angle when the test vehicle transitions from accelerating to hard braking. The vehicle model, incorporating Eq. \eqref{eqn:wt_model}, is then used to simulate the pitch response given similar acceleration inputs (orange line, Fig. \ref{fig:wt_calibration}). The parameter $k=3.01$ is selected such that the vehicle model simulation approximates the behavior in the test data.

\subsection{Lateral Brake Balance} \label{sec: lateral_bb}

In addition to longitudinal weight transfer, lateral weight transfer can also change the normal force on each tire. This is particularly important to consider in corners and during trail braking. Previous work has shown the importance of utilizing lateral brake balance in high performance driving \cite{SubositsThesis}. Here, in order to balance computational complexity of the NMPC layer, we propose a novel hierarchical control approach for calculating lateral brake balance. The allocation of brake torques on the right and left side of each axle is calculated outside the NMPC in a low level controller and is based upon the static lateral load transfer, as given by the reference trajectory. For the right and left side of a single axle, this is given respectively by:

\begin{align}
    \tau_{brake,r} &= \tau_{brake,axle} \left( \frac{1}{2} + \frac{a_y h_{cg}}{t_{width}g}\right) \\
    \tau_{brake,l} &= \tau_{brake,axle} \left( \frac{1}{2} - \frac{a_y h_{cg}}{t_{width}g}\right)
\end{align}

Where $\tau_{brake,axle}$ is the total braking force commanded by the NMPC to either the front or rear axle, and $\tau_{brake,r}$ and $\tau_{brake,l}$ are the individual brake torques for the right and left side of the axle. Additionally, $g$ is the acceleration due to gravity, $t_{width}$ is the vehicle track width, and $a_y$ is the lateral acceleration. Because the low level controller commands different brake torques to the right and left side of the vehicle, an additional yaw moment is created, shown below:

\begin{equation}
\begin{split}
        \tau_{bb} = \frac{-(\tau_{brake,fl}- \tau_{brake,fr})}{r_w}\cos(\delta)(t_{width}/2) - \frac{(\tau_{brake,rl}- \tau_{brake,rr})}{r_w}(t_{width}/2)
\end{split}
\end{equation}

where $\tau_{brake,fr}$ and $\tau_{brake,fl}$ are the front right and left brake torques, and $\tau_{brake,rr}$ and $\tau_{brake,rl}$ are rear right and left brake torques. The term $\tau_{bb}$ is accounted for in the model predictive control layer, and the importance of doing so is discussed in Sec. \ref{lateral_brake}.

\subsection{Gear Change Algorithm}
Optimizing gear choice inside the NMPC is considered out of scope for this paper. Gear changes are computed outside of the NMPC in a low level controller based on the reference path. The engine torque from the NMPC solution is converted to an overall drive force, and then matched considering the current gear of the vehicle.

\section{MPC Formulation}
Two optimal control problems (OCPs) are used, one for generating the reference trajectory which optimizes for the entire track length, and a fixed horizon MPC used online.  The reference trajectory is generated using a similar formulation as the online controller, but the focus of this paper is on the online controller. The MPC formulation is given in general form as:

\begin{equation}
\begin{aligned}
    \min{J} \\
    s.t.\quad & x_{k+1} = f(x_k,u_k) \;\;\qquad \forall k \in [1,N]\\
    \quad & g(x_k,u_k)=0 \;\qquad\qquad \forall k \in [1,N]\\
    \quad & h(x_k,u_k) \le 0 \;\qquad\qquad \forall k \in [1,N]\\
    \quad & x_{min} \leq x_k \leq x_{max} \qquad \forall k \in [1,N]\\
    \quad & u_{min} \leq u_k \leq u_{max} \qquad \forall k \in [1,N]\\
    \quad & x_0 = x_{lookahead}
\end{aligned}
\end{equation}

With $J$ being the cost, $x$ the state vector, and $u$ the input vector. $x_{min}$ and $x_{max}$ are the minimum and maximum values for the state vector, respectively.  $u_{min}$ and $u_{max}$ are defined in the same way for the inputs.  Lastly, the initial state, $x_0$, is constrained to be equal to the state vector $x_{lookahead}$, which is constructed using the method in Section \ref{sec:x0}. 

To efficiently encode the reference path and trajectory states, the dynamics are represented in spatial terms along the curvilinear coordinate system relative to the reference trajectory:

\begin{equation}
    \frac{dx}{ds} = \frac{1}{\dot{s}} \frac{dx}{dt}
\end{equation}
\begin{equation}
    \dot{s} = \frac{V\cos \Delta\phi}{1 - \kappa_{ref} e}
\end{equation}

The vehicle dynamics are discretized using a second-order implicit Runge Kutta method, which has been shown to balance accuracy with computational effort \cite{BrownRK2}. In order to have increased integration accuracy in the first part of the horizon but still maintain an appropriately long look ahead distance, the first 5 points of the horizon have a step length of $ds=3m$ and the remaining horizon points have a step length of $ds=7m$. We use 20 points in the NMPC horizon giving a lookahead distance of $120m$.

\subsection{Cost}
For the online control formulation, the cost function is given as
\begin{equation}
\begin{split}
    J = J_{s_N} + \sum_{i=0}^{N} &(J_{e_i} + J_{t_i} + J_{\alpha_i} + J_{x_i} +\\ 
    & J_{F_i} + J_{\ddot{u}_i} + J_{u})\cdot ds_i
\end{split}
\end{equation}

Where $N$ is the horizon length and $ds_i$ is the path distance step length.  The running cost consists of several terms penalizing the state, deviation from the reference trajectory, and control effort, weighted by the step lengths at each step. 

\subsubsection{State Bounds Cost}
To prevent infeasiblity due to constraint violations at the first stage of the NMPC problem, which is propagated from measured vehicle states, we chose to implement the track bound and sideslip violation as soft constraints. The state bound cost imposes a slack constraint on track bound violation and exceeding a specified maximum vehicle sideslip.  The components of this cost are only active if the maximum or minimum values are exceeded.  When exceeded, this cost is given as

\begin{equation}
    J_{e_i}= w_{tb}(e_i-e_{min_i,max_i})^2 + w_{\beta}(\beta_i-\beta_{min,max})^2
\end{equation}

Where $w_{tb}$ is a large weight on violating track bounds, and $w_{\beta}$ is a large weight on exceeding the prescribed sideslip range.

\subsubsection{Time and Tracking Cost}
This tracking cost penalizes the lateral error from the reference trajectory, as well as time, $t=(ds_i/\dot{s_i})$, accumulated over the horizon.  This is given as

\begin{equation}
    J_{t_i} = w_e {e_i^2} + w_t {t}
\end{equation}

Where $w_e$ is a weight on the lateral error, and $w_t$ is a weight on time.

\subsubsection{Front Tire Sideslip}
To aid convergence, a small regularization cost is imposed on the front tire sideslip to avoid zero gradients at tire force saturation, inspired by \cite{LaurenseThesis}.  This is given as

\begin{equation}
    J_{\alpha_i} = w_{\alpha}{\alpha_{f_i}^2}
\end{equation}

With the weight $w_\alpha$ weighting the sideslip.

\subsubsection{State Regularization Cost}
The state regularization cost imposes a small cost penalizing deviation from the reference velocity and brake torques. This is given as

\begin{equation}
\begin{split}
    J_{x_i} = w_V{(V_i-V_{ref_i})^2} +\\
    w_{\tau_{brake_f}}{(\tau_{brake_f,i}-\tau_{brake_f,ref,i})^2} + \\
    w_{\tau_{brake_r}}{(\tau_{brake_r,i}-\tau_{brake_r,ref,i})^2}
\end{split}
\end{equation}

Where $w_V$ is the velocity weight, and $w_{\tau_{brake_f}}$ and $w_{\tau_{brake_r}}$ are front and rear brake torque weights, respectively. $V_{ref}$ is the reference velocity, while $\tau_{brake_f,ref}$ and $\tau_{brake_r,ref}$ are the reference front and rear brake torques, respectively. When testing, $V_{ref}$ is approximately scaled for consistency with the imposed force circle limit cost in Section \ref{sec:force_circle}.

\subsubsection{Force Circle Cost}
\label{sec:force_circle}
This cost penalizes exceeding a designated maximum fraction of the friction circle at the lumped front and rear tires, accounting for longitudinal load transfer. This is encoded as a soft constraint, so that the vehicle can use additional force if necessary, e.g. to abide by the road bounds cost. This also prevents infeasibility due to initial conditions. When the tire force exceeds the friction circle, this is given as

\begin{equation}\label{eq: force_circle}
\begin{split}
    J_{F_i} = w_F\bigg[\bigg(\frac{F_{xf_i}^2+F_{yf_i}^2}{(\mu_f F_{zf_i})^2}-(\mu_{lim})^2\bigg)^2 +\\  \bigg(\frac{F_{xr_i}^2+F_{yr_i}^2}{(\mu_r F_{zr_i})^2}-(\mu_{lim})^2\bigg)^2\bigg]
\end{split}
\end{equation}

Where $w_F$ is a weight and $\mu_{lim}$ is the designated maximum fraction of the estimated friction. $F_z$ is the load on each tire accounting for longitudinal load transfer and topology, which directly impacts the force potential at each tire. 

\subsubsection{Input Acceleration Cost}
The input acceleration cost penalizes the engine torque and steering angle acceleration to promote smooth inputs:

\begin{equation}
    J_{\ddot u_i} = w_{\ddot{\delta}} \ddot{\delta_i}^2 + w_{\ddot{\tau}} \ddot{\tau_i}^2
\end{equation}
With $w_{\ddot{\delta}}$ and $w_{\ddot{\tau}}$ being the corresponding weights.

\subsubsection{Input Cost}
The input cost applies a small regularization to the reference brake torque rate:

\begin{equation}
    J_{u} = w_{\dot{\tau}}(\dot{\tau}_{brake,r,i}-\dot{\tau}_{brake,r,ref,i})^2 + w_{\dot{\tau}}(\dot{\tau}_{brake,f,i}-\dot{\tau}_{brake,f,ref,i})^2
\end{equation}
Where $w_{\dot{\tau}}$ is the weight.

\subsubsection{Terminal Stability Cost} \label{sec:error_stability_cost}
The terminal stability cost encodes sideslip and error stability by encouraging first order dynamics for path error and sideslip at the terminal state. Specifically:

\begin{equation}
    J_{s_N} = w_{\dot{\beta}_N}ds_N(\dot{\beta}_N+k_{\dot{\beta}}\beta_N)^2 + w_{\dot{e}}ds_N(\dot{e}_N+k_{\dot{e}} e_N)^2
\end{equation}

With $w_{\dot{\beta}}$ being a weight on sideslip rate, and $w_{\dot{e}}$ being a weight on the lateral error rate.  $k_{\dot{\beta}}$ and $k_{\dot{e}}$ are constants.

\subsection{Constraints}
\subsubsection{Initial State Constraints}
\label{sec:x0}
The initial state of the NMPC problem, $x_0$, is constrained to be equal to the lookahead state, $x_{lookahead}$. To construct $x_{lookahead}$, the current state of the vehicle is integrated forward by $t_{lookahead}=50ms$ to account for the expected solve time of the controller. 

For the NMPC states which correspond to the input states, namely $\delta$, $\tau$, $\tau_{brake,f}$, and $\tau_{brake,r}$, the corresponding terms of $x_{lookahead}$ are calculated by starting with the value inside the current NMPC solution, and integrating forward by $t_{lookahead}$ using the inputs $\dot{\delta}, \dot{\tau}, \dot{\tau}_{brake,f}, \dot{\tau}_{brake,r}$ also from the current NMPC solution. This is done to achieve smooth inputs from the NMPC. This procedure is also used for states where there is not a measurement available, $dF_z$, or the available measurement is noisy, $\omega_r$.

For states where measurements are available, namely $r, V, \beta, e$, and $\Delta\phi$, the corresponding terms of $x_{lookahead}$ are calculated by starting with the current measurements and integrating forward by $t_{lookahead}$ using the vehicle model in Section 2.1, and the input roll-outs from above.

\subsubsection{Actuation Constraints}
Maximum and minimum bounds are imposed on the inputs and states to maintain consistency with the vehicle’s physical limitations, e.g. steering range, steering motor power, maximum steering slew rate, engine torque and power limits, and maximum engine/brake torque slew rates.  This is given as

\begin{equation}
    \begin{bmatrix}
    \delta_{min} \\
    \dot{\delta}_{min} \\
    \ddot{\delta}_{min} \\
    \omega_{r,min} \\
    \tau_{min} \\
    \tau_{brake, min} \\
    \tau_{brake,min} \\
    \dot{\tau}_{brake, min} \\
    \dot{\tau}_{brake,min} \\
    \end{bmatrix}
    \leq
    \begin{bmatrix}
    \delta \\
    \dot{\delta} \\
    \ddot{\delta} \\
    \omega_{r} \\
    \tau \\
    \tau_{brake, f} \\
    \tau_{brake,r} \\
    \dot{\tau}_{brake, f} \\
    \dot{\tau}_{brake,r} \\
    \end{bmatrix}
    \leq
    \begin{bmatrix}
    \delta_{max} \\
    \dot{\delta}_{max} \\
    \ddot{\delta}_{max} \\
    \omega_{r,max} \\
    \tau_{max} \\
    \tau_{brake, max} \\
    \tau_{brake,max} \\
    \dot{\tau}_{brake, max} \\
    \dot{\tau}_{brake,max} \\
    \end{bmatrix}
\end{equation}

\subsubsection{Dynamic Longitudinal Brake Balance}
With load being transferred longitudinally between the front and rear axles and laterally during acceleration and cornering, the load on each tire varies throughout operation. For example, as load is shifted forward during braking, the front tires have more capability to generate forces due to the increased load, and concomitantly, the rear tire has less.  As such, a dynamic brake balance that can allocate brake torques among each tire independently is important to exploit the full capabilities of the vehicle.  To allow the NMPC to optimally allocate these forces, longitudinal brake torque on the front and rear axle are treated as separate states. Because the static weight distribution of the test vehicle is biased to the front, and load transfers forward during braking, there are very few cases in which it is advantageous to brake more in the rear than in the front. It was found empirically that constraining the front brake torque to be larger in magnitude than the rear brake torque aided convergence times, without limiting the practical performance of the controller. With the convention that brake torques are always negative, this constraint is expressed as:

\begin{equation}
    \tau_{brake,r}>\tau_{brake,f}
\end{equation}

\begin{figure}
    \centering
    \includegraphics[width=0.6\textwidth, bb=0 0 960 960]{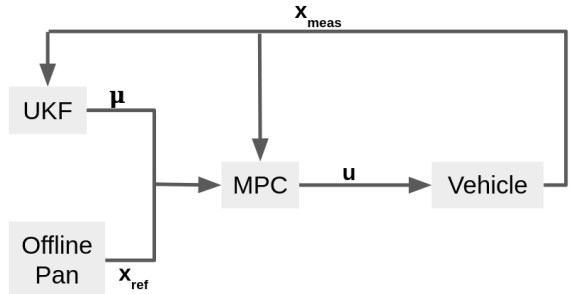}
    \caption{High-level diagram of the adaptive MPC formulation.}
    \label{fig:high_level}
\end{figure}

\subsection{Solver Implementation}
The OCPs are implemented using the CasADi auto-differentiation and code-generation toolbox\cite{Andersson2019} and solved with the interior point method implemented with IPOPT \cite{IPOPT}. To improve convergence time, and to keep the solutions within a similar local minimum attraction basin, the initial guess is set to the previous converged optimal solution; or the offline reference if no previous optimal solution exists. The solver is constrained to a maximum of 50 iterations, preventing runaway computation. The command sent to the vehicle is interpolated from the most recent optimal solution based upon the current path distance, $s$.

\section{Unscented Kalman Filter}
To estimate the coefficient of friction, we build on the UKF approach in \cite{DALLAS202011,DALLASIFAC, WurtsACC}, as this has been shown to be of suitable balance between efficiency and accuracy and can better approximate nonlinear transformations than an Extended Kalman filter \cite{Simon}. By accounting for longitudinal and lateral force coupling, and the impact of load transfer and topology, we robustly deploy this technique from simulations to real-world experiments.  Briefly, the UKF predictions are performed using the bicycle model with lateral yaw perturbation, described in Eq. \eqref{eq:bicycle}; however the prediction model only contains five states; the yaw rate, velocity, sideslip, and front and rear friction (represented with trivial dynamics).  The UKF correction step is based upon measurements of the yaw rate, velocity, and sideslip. The UKF runs at 62.5 Hz and the MPC bicycle model is updated with the current friction estimate at each call. A high level diagram describing the integration is shown in Fig {\ref{fig:high_level}}.

\subsection{Tuning}
The UKF is automatically tuned for the process noise covariance matrix and initial friction variance.  The process optimizes the following function

\begin{equation}
\begin{aligned}
      \min{J = \sum_{i=0}^{N}{(y_{pred,i}-y_{meas,i})^2 + (\mu_{pred,i}-\mu_{true,i})^2}} \\
    s.t.\quad Q_{min} \le Q \le Q_{max} \\
    \quad \sigma^2_{min} \le \sigma^2 \le \sigma^2_{max} \\
\end{aligned}
\end{equation}

where $y_{pred}$ is the normalized predictions of the state vector consisting of yaw rate, velocity, and sideslip, and $y_{meas}$ are the measurements. $\mu_{pred,i}$ is the UKF prediction of the front and rear frictions at each index, while $\mu_{true,i}$ is the believed truth for the friction. $Q$ is the process noise covariance matrix, and $Q_{min}$ and $Q_{max}$ represent the upper and lower bounds respectively.  Bounds are also placed on the initial friction variance, $\sigma^2$. Inside each iteration of the optimization, two steps occur.  First, the UKF is run to obtain point-wise estimates of friction for the given process noise.  Second, open loop predictions, parameterized by the UKF estimates, are performed over the entire data set.  The cost is then evaluated using these state predictions and friction estimates, and then the process noise and initial friction variance are updated and the next iteration begins. 

To prevent overfitting, the dataset is structured as two back-to-back laps of a representative test track with three different initial conditions; for $N=6$ laps (approximately 12 minutes) in total. This consists of two laps with a high initial guess, two with a low initial guess, and two with the nominal initial guess.  The cost function is designed to determine the process noise covariance matrix that minimizes the open loop prediction error as compared to measurements.  The optimization problem is solved using the L-BFGS implementation in the SciPy toolbox. \cite{2020SciPy-NMeth}.

\section{Experimental Vehicle} \label{sec: exp_veh}

\begin{figure}
    \centering
    \includegraphics[scale=0.3, bb=0 0 960 960]{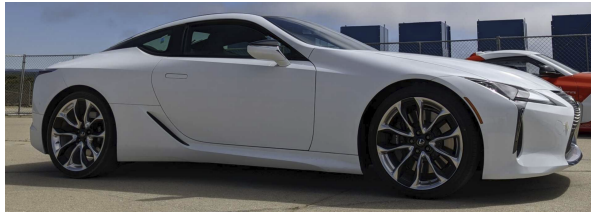}
    \caption{LC 500 experimental vehicle.}
    \label{fig:LC500}
\end{figure}

Experiments are performed on a 2019 Lexus LC 500, shown in Fig. \ref{fig:LC500}. The powertrain, drivetrain, and suspension are not modified. Autonomous control is achieved by communicating with the pre-existing driver assistance features and hardware, which required extensive modification. The vehicle is equipped with an Oxford Technical Systems (OxTS) RT3000 v3 RTK-GPS/IMU system with dual antennas for localization and state estimation. The MPC computation is performed by a RAVE ATC8110-F ruggedized computer running Ubuntu Linux. Low level control and communication is handled by a dSpace MicroAutoboxII (DS1401). The OxTS data and all actuator commands are communicated via CAN, and communication between the Linux computer and the MicroAutoBox is done via UDP. All experiments were performed on a closed course.

\section{Results and Discussion}
\begin{figure}
\centering
\begin{subfigure}[b]{0.45\textwidth}
    \centering
    \includegraphics[width=\textwidth]{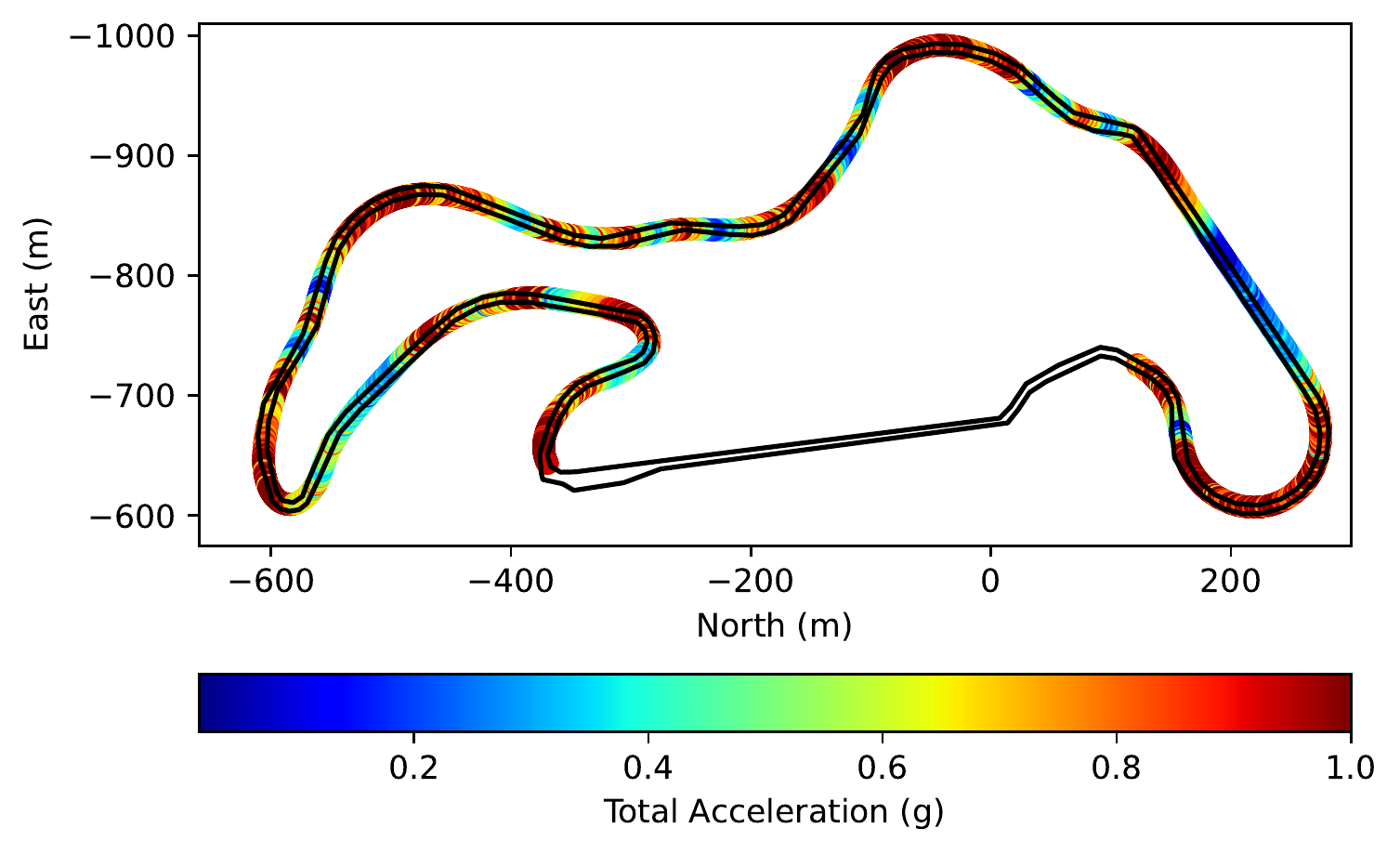}
    \caption{Thunderhill West 2-mile track with total acceleration overlayed.}
    \label{fig:twoMile}
\end{subfigure}
\hfill
\begin{subfigure}[b]{0.45\textwidth}
    \centering
    \includegraphics[width=\textwidth]{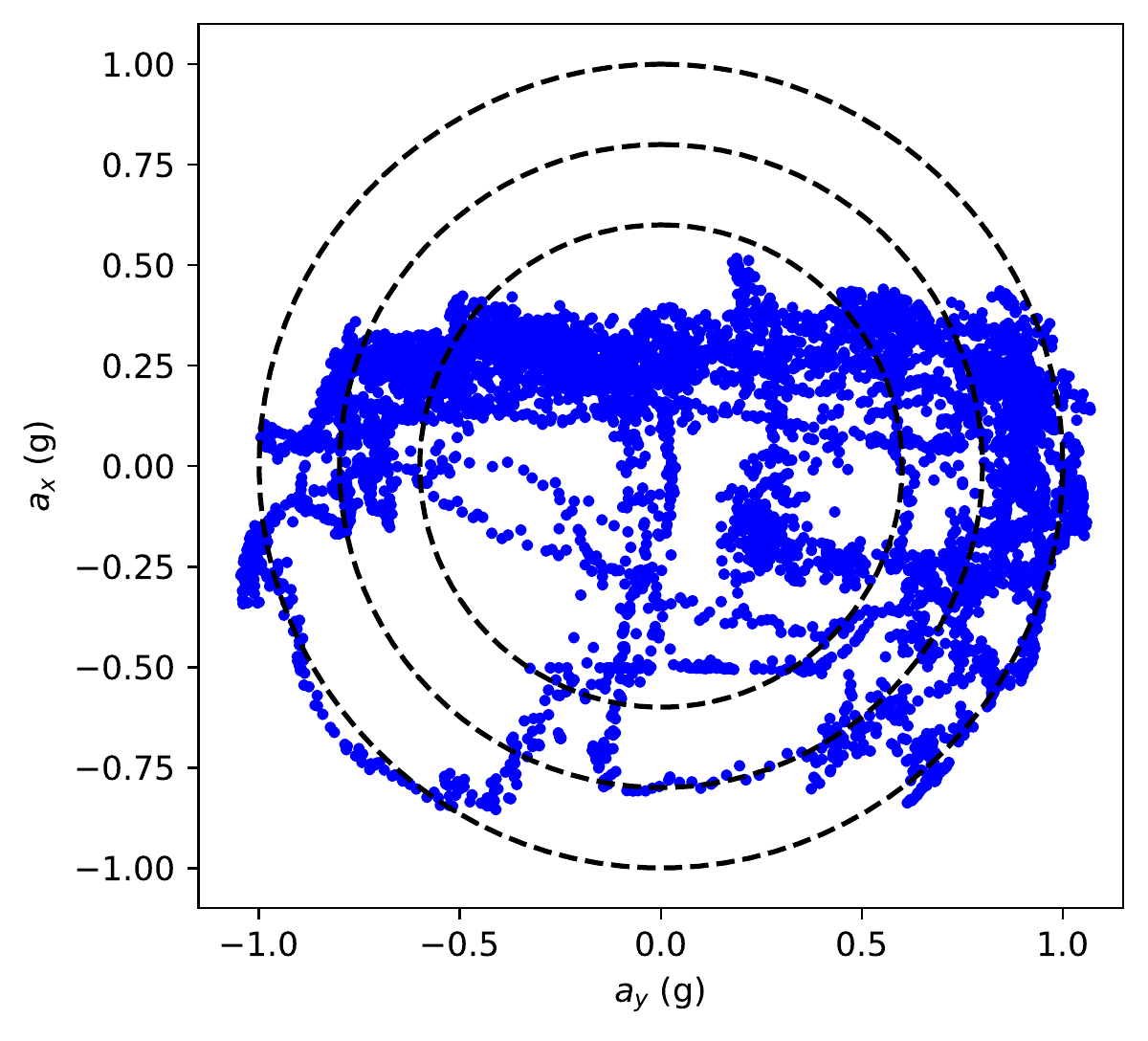}
    \caption{GG acceleration plot. Dashed circles represent 0.6g, 0.8g, and 1.0g.}
    \label{fig:twoMilegg}
\end{subfigure}
\caption{Thunderhill track test with friction limit of 0.95$\cdot\mu$}
\end{figure}
The NMPC formulation was tested with the experimental vehicle of Sec. \ref{sec: exp_veh} on the Thunderhill West 2-mile track.  This section provides the results of the integrated algorithm on a racetrack. Furthermore, comparative experiments that examine the importance of online adaptation, dynamic brake balance, and yaw moment created from lateral brake proportioning are performed.

\subsection{Integrated Approach on Race Track}
To evaluate the overall integrated approach including online friction estimation, the controller was tested on the Thunderhill West racetrack with the force friction circle limit (Sec. \ref{sec:force_circle}) set to $0.95\cdot\mu$. This 2-mile long course includes sections with significant grade and bank, and several typical race track features including chicanes and hairpin turns. The straight in the bottom of Fig. \ref{fig:twoMile} is not tested autonomously due to safety concerns from a concrete wall immediately adjacent to the track in this section. Fig. \ref{fig:twoMile} shows the measured total acceleration (g-forces) of the vehicle as it autonomously drives the track. In braking zones and turns, the vehicle regularly operates close to $1g$ of acceleration, and the engine is often at full power during corner exits. On multiple segments, the vehicle reaches speeds of 39 $m/s$ and is immediately followed by sharp turns that require a drastic deceleration. This requires the enhanced braking potential achieved through the longitudinal and lateral brake distributions. This is depicted in the GG diagram in Fig. \ref{fig:twoMilegg} which highlights the vehicle consistently operating near its maximum capabilities, both in pure lateral cornering and also in combined trail braking and acceleration. 

\begin{figure}
    \centering
    \includegraphics[scale = 0.75]{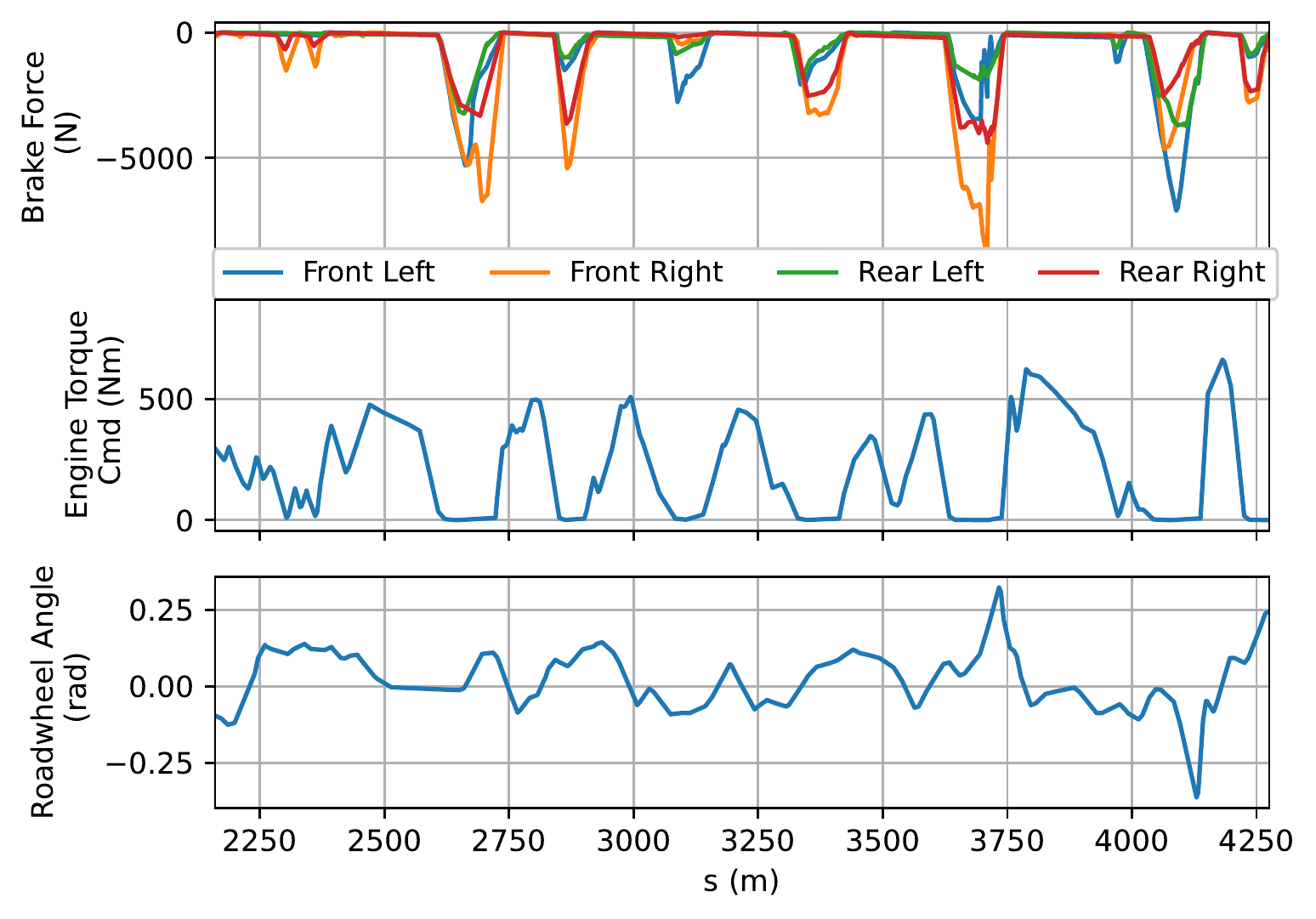}
    \caption{Experiments control inputs for brake force (top) for front left wheel (blue), front right (orange), rear left (green), and rear right (red), engine torque (middle), and road wheel angle (bottom).}
    \label{fig: TwoMileInputs}
\end{figure}

\begin{figure}
    \centering
    \includegraphics[scale = 0.75]{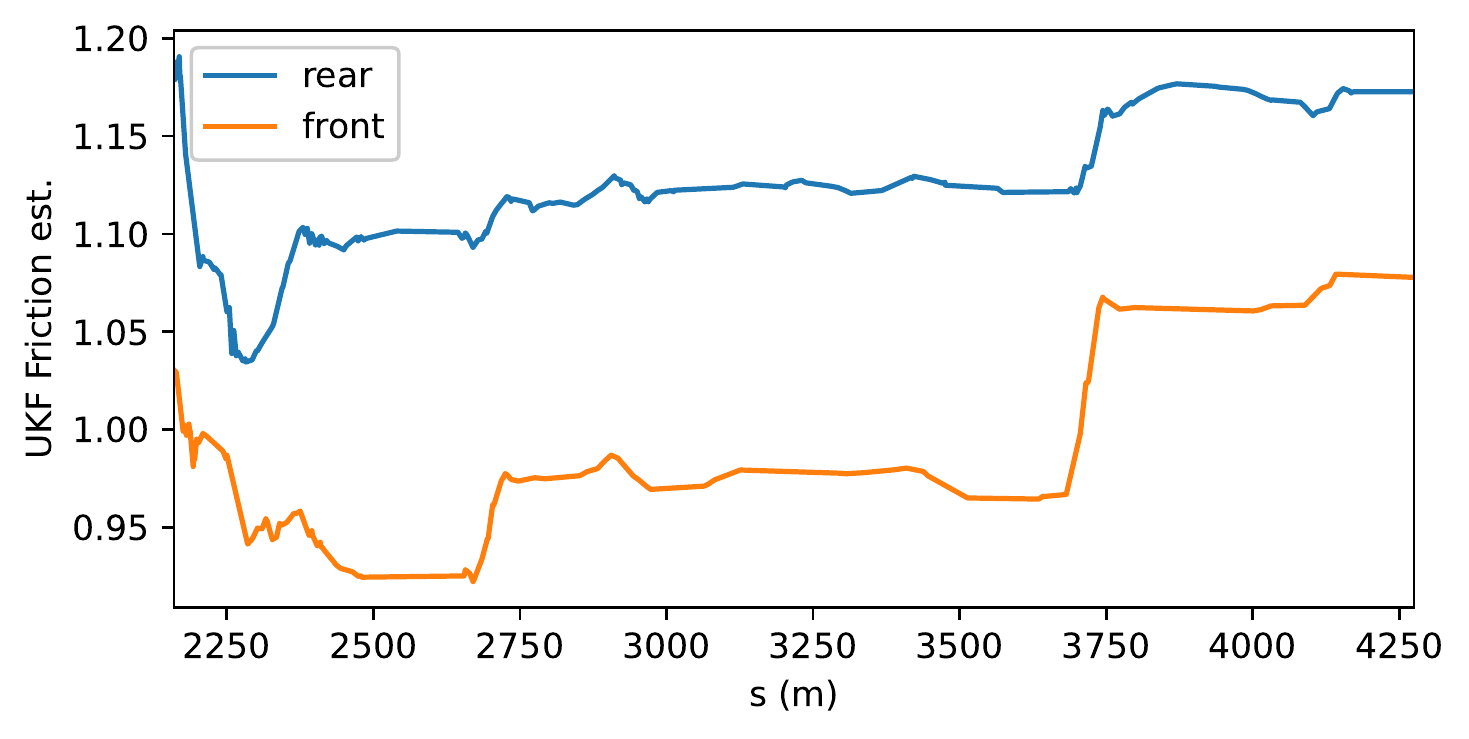}
    \caption{UKF friction estimate for lap around Thunderhill West.}
    \label{fig:ukf_hotlap_est}
\end{figure}

The high performance of this trail braking behavior is enabled by the dynamic lateral and longitudinal brake proportioning. This is displayed in the top plot of Fig. \ref{fig: TwoMileInputs}, where the commanded brake torque is different for each wheel to account for the lateral and longitudinal weight transfer. The engine torque commands and road wheel angle are shown in the middle and bottom of Fig. \ref{fig: TwoMileInputs} respectively, and remain smooth throughout this challenging trajectory.

\begin{figure}
    \centering
    \includegraphics[scale = 0.75]{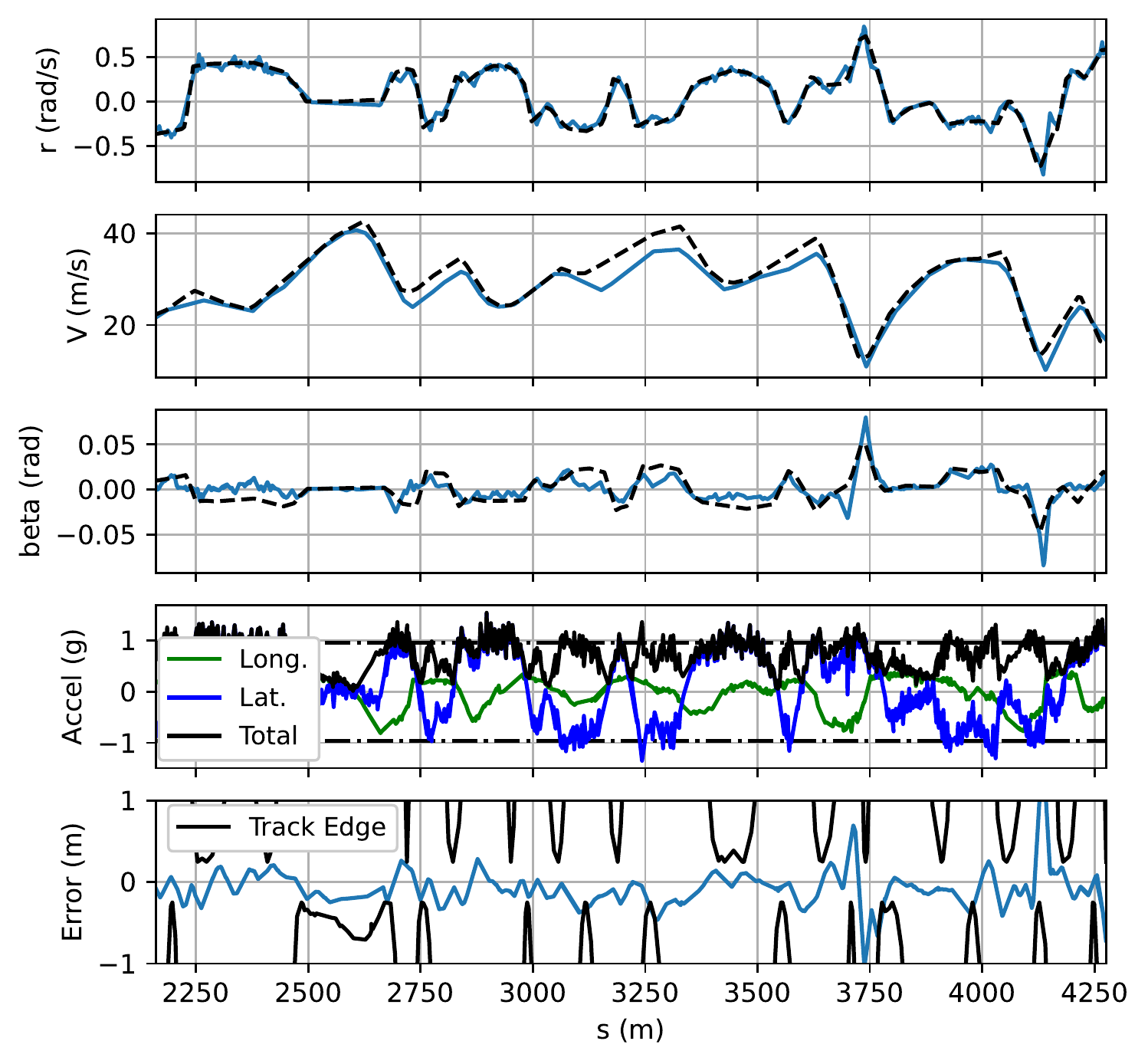}
    \caption{Vehicle state trace for Thunderhill West at a 0.95$\cdot\mu$. Reference trajectory for $r$, $V$, and $\beta$ is shown as a dashed black line. The acceleration plot has the friction limit of $0.95\cdot\mu$ as a dashed dotted line. The lateral error plot has the track edge as a solid black line.}
    \label{fig:th_state}
\end{figure}

The acceleration data in Fig. \ref{fig:th_state} and Fig. \ref{fig:twoMile}, is obtained from the GNSS-IMU unit and is filtered using the Scipy \cite{2020SciPy-NMeth} implementation of a bidirectional 4th order digital low pass Butterworth filter with a natural frequency of 9 Hertz. The second plot from the bottom in Fig. \ref{fig:th_state} shows the lateral, longitudinal, and magnitude of total acceleration. The black dash dotted line at 0.95$\cdot g$ represents the acceleration target corresponding to a friction limit of 0.95$\cdot\mu$. This figure showcases the trail braking ability of this controller. Particularly between $s=3250$ and $s=3400$, the controller starts with hard braking and large longitudinal acceleration, and smoothly releases the brakes while increasing lateral acceleration, thereby keeping the total acceleration at nearly constant magnitude through the turn. Similar behavior can be seen throughout, e.g. this is also demonstrated between $s=2500$ and $s=2750$. 

An inherent and positive attribute of NMPC controllers is the ability to re-plan locally optimal control inputs when operating far from the reference trajectory. With this in mind, we intentionally let the NMPC framework in this work deviate from the reference path in order to minimize time around the race track even if the test conditions differ from the reference path conditions. This characteristic is displayed in the bottom plot of Fig. \ref{fig:th_state} which shows the error from the reference path as well as the track edges. The controller regularly deviates from the reference path, adapting to changing track conditions and the current state evolution to minimize time. For example at $s=2500m$ the vehicle pushes close to the track edge in order to widen the corner entry of turn 3, then crosses the reference path and pushes to the other side of the track at $s=2725$ towards the apex of the corner. 

Operating away from the reference path becomes a necessary feature to fully take advantage of the UKF friction estimation, which will often estimate a different friction value than the reference path conditions. As seen in Fig. \ref{fig:ukf_hotlap_est}, the friction values estimated by the UKF are slightly lower in the middle of the track from $s\approx2500$ to $s\approx3300$. The estimated front friction drops as low as $\mu_{front}=0.93$ compared to the reference trajectory value of $\mu_{front}=1.004$. Various unmodeled factors can create this result including uneven heating of the track, road surface changes, or inaccurate topology information. As a result, the NMPC controller reduces velocity in this section relative to the reference, which is planned with a nominal friction value for the entire track. Fig. \ref{fig:th_state} shows the reduced velocity in this middle section, which returns nearer to the reference as the UKF friction estimate returns closer to the nominal friction value. This demonstrates the ability of this controller to fully use but not exceed the available friction, even if the test conditions differ from the reference conditions. 

Overall, the vehicle was able to exhibit precise control at the limits of handling at a friction limit of $0.95\cdot\mu$. Importantly, throughout this experiment, the mean solve time was 31.3 milliseconds with a standard deviation of 6.7 milliseconds, and a mean number of iterations of 15.7 with a standard deviation of 3.1, demonstrating the ability for real time application.  In the following subsections, comparative tests highlight the importance of several key components of this formulation in achieving this performance.

\subsection{Weight Transfer Modeling and Brake Balance} \label{wt_brake_balance}

The most direct impact of including weight transfer in the NMPC vehicle model is the ability to dynamically change the brake balance while braking. As the ideal braking force for the front and rear axle is dependent on the normal force at each axle, the longitudinal weight transfer during braking can have a large impact on the optimal braking force and brake balance. At the beginning of a flat braking zone, the weight distribution of the car is approximately equal to the static weight distribution, making $\tau_{front} / \tau_{rear} = 1.13$ the optimal brake balance for the test vehicle at this point. As the vehicle brakes, more load is transferred to the front axle thereby reducing the capability of generating force at the rear axle, shifting the optimal brake balance to bias the front axle.

\begin{figure}
    \centering
    \includegraphics[scale = 0.75]{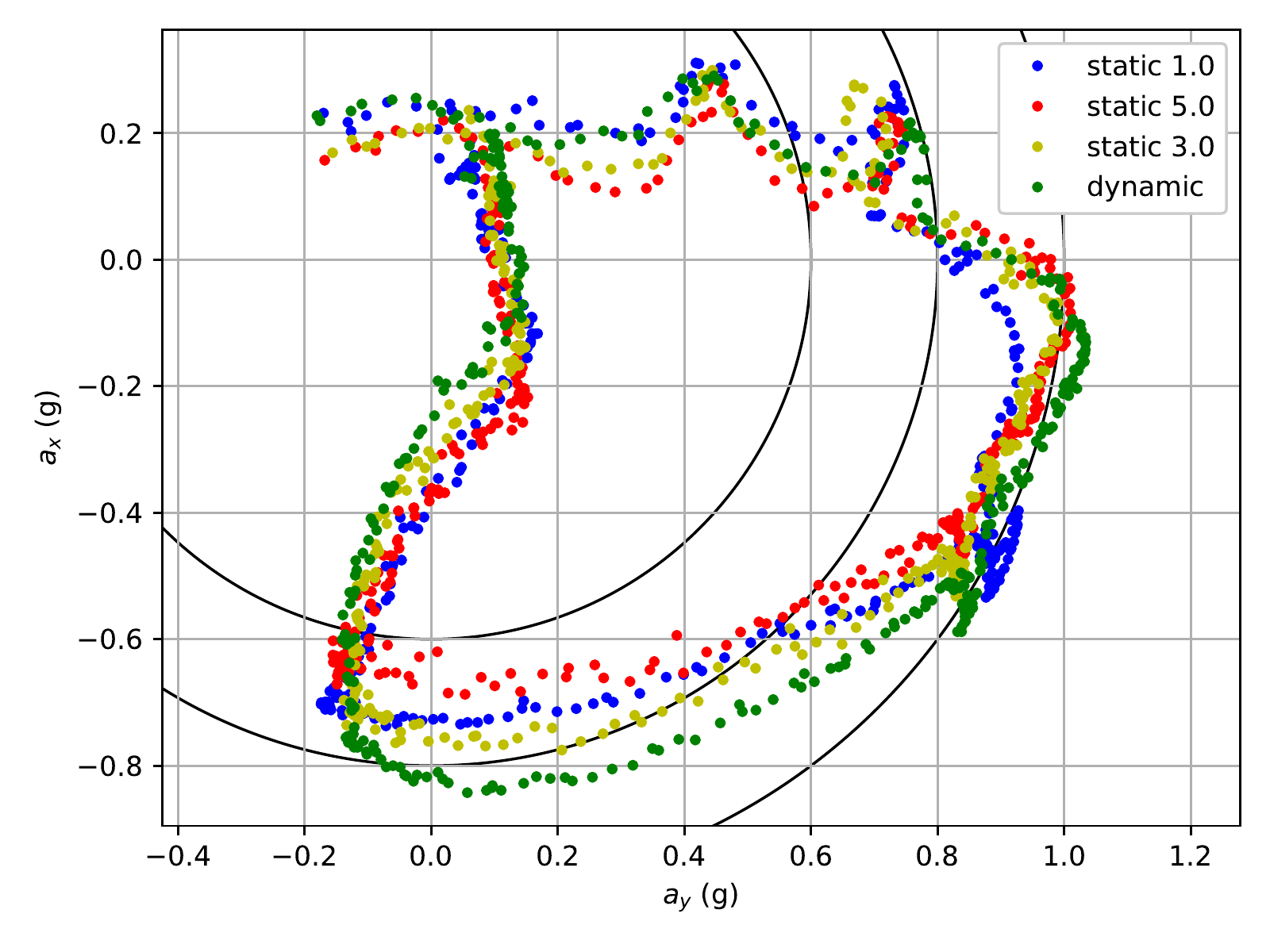}
    \caption{Acceleration for turn 3 at Thunderhill West showing increased performance with weight transfer modeling and dynamic brake balance (green) as compared to static distributions of 1.0 (blue), 3.0 (yellow) and 5.0 (red).}
    \label{fig:wt_gg}
\end{figure}

\begin{figure}
    \centering
    \includegraphics[scale = 0.75]{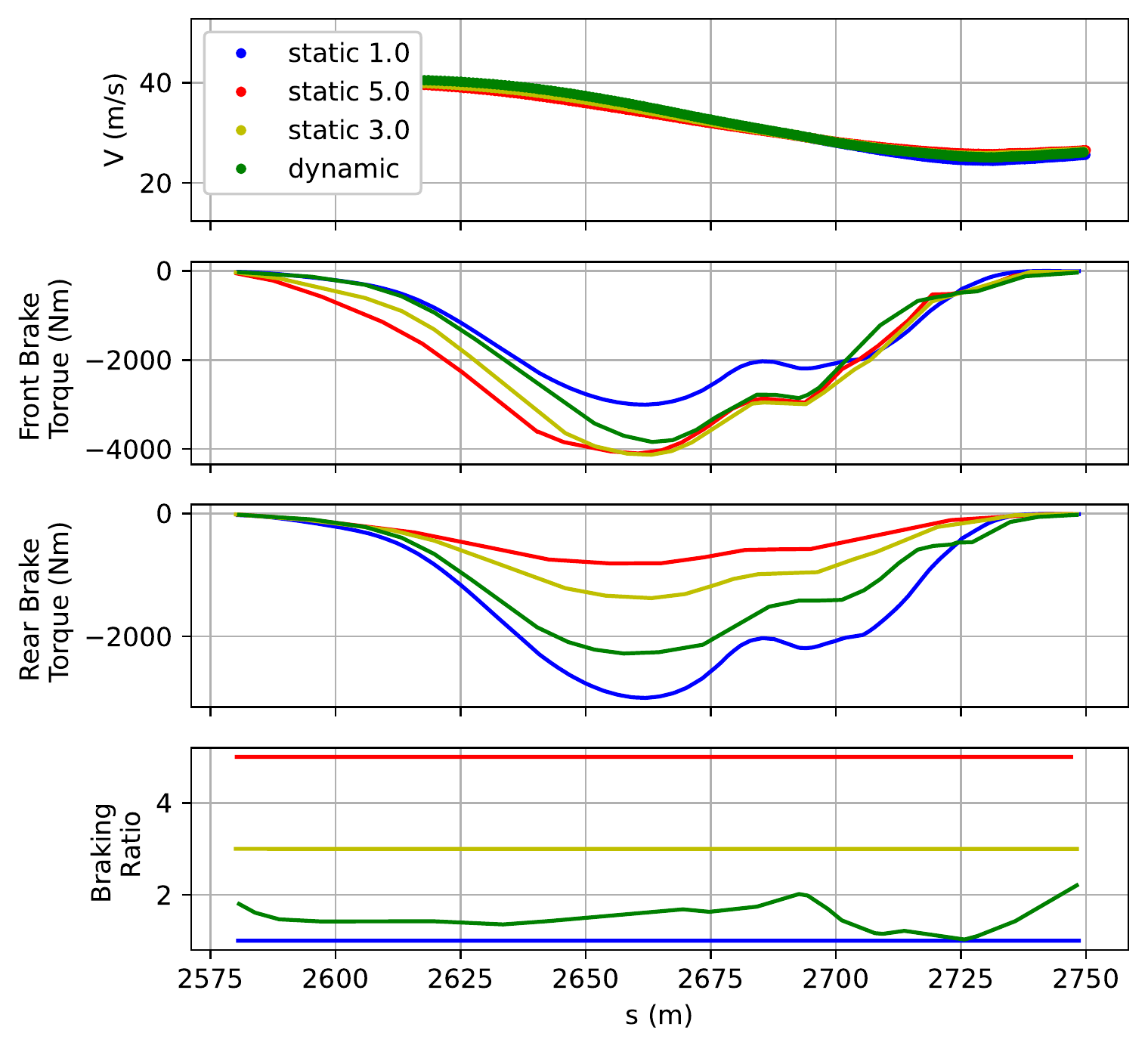}
    \caption{(a) velocity plot, (b) front brake torque, (c) rear brake torque, and (d) brake ratio for turn 3 at Thunderhill West showing increased performance with weight transfer modeling and dynamic brake balance (green) as compared to static distributions of 1.0 (blue), 3.0 (yellow) and 5.0 (red).}
    \label{fig:wt_bb}
\end{figure}

Demonstrating these effects, Fig. \ref{fig:wt_gg} and \ref{fig:wt_bb} shows the results of four autonomous tests where the vehicle starts braking and entering a turn. In one test the longitudinal weight transfer is modeled inside the NMPC and the dynamic brake balance is applied. In the other tests, weight transfer is not modeled, and instead a constant brake balance ratio of $\tau_{front} / \tau_{rear}=1$, $3$, and $5$ are included as a slack constraint inside the NMPC.  For these tests, the test vehicle autonomously drives through turn 3 of the Thunderhill West track and is aiming to use up to 90\% of the available tire friction. Turn 3 of Thunderhill West has a fast and slightly downhill approach, a flat braking zone, followed by a left turn. In all tests, the vehicle approaches turn 3 at a velocity of approximately 40 meters per second and slows to around 25 meters per second at the apex of the turn.  

Fig. \ref{fig:wt_gg} shows the G-G trace, the longitudinal vs. lateral acceleration expressed as a ratio of $g$. Without modeling weight transfer, the vehicle is only able to achieve longitudinal acceleration of about 0.7g for $\tau_{front} / \tau_{rear}=5$, 0.77g for $\tau_{front} / \tau_{rear}=3$, and 0.74g for $\tau_{front} / \tau_{rear}=1$ and does not utilize the imposed limit of maximum braking force. When weight transfer is modeled and dynamic brake balance utilized, the vehicle is able to achieve 0.82g of longitudinal acceleration. As the vehicle turns, a larger trace is observed in the trail braking section for the dynamic brake balance as compared to the static brake balance, as the vehicle better utilizes the available force.  

Fig.  \ref{fig:wt_bb} depicts the measured vehicle speed through the turn. Despite starting at similar speeds, the approach with dynamic brake balance is able to carry more speed through the trail braking and cornering sections, while also achieving similar exit speed. For the dynamic brake balance, the front brakes are delayed as compared to the static distribution (middle plot) and are released sooner since the vehicle is able to achieve higher magnitude acceleration from braking.  This leads to the vehicle braking later and carrying more speed early in the turn, but losing speed rapidly in the braking zone (top). Fig. \ref{fig:wt_bb} (bottom) shows the brake balance during the test. The NMPC with weight transfer starts braking with a brake balance of $\tau_{front} / \tau_{rear} = 1.4$, factoring in the vehicle acceleration and topology that can cause the higher distribution than the static value of 1.13. As load is transferred during braking, the brake distribution increases to $\tau_{front} / \tau_{rear} = 2.0$ before easing off as the brakes are released. This indicates that the dynamic brake balance allows the NMPC to utilize the available rear braking force at the start of braking, and also utilize the increased available braking force at the front axle as load is transferred to the front.

\subsection{Lateral Brake Distribution Model} \label{lateral_brake}
\begin{figure}
    \centering
    \includegraphics[scale = 0.6]{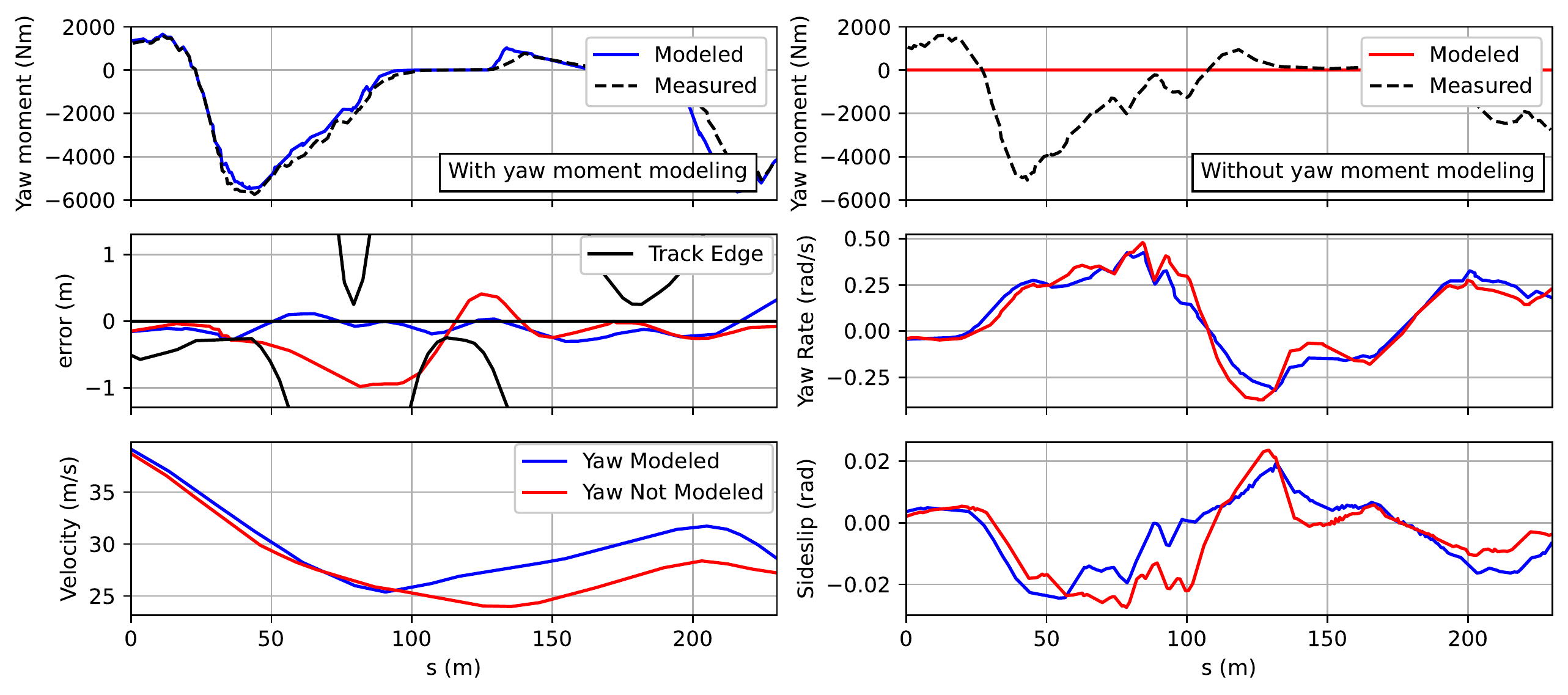}
    \caption{Performance of controller with and without yaw moment model.}
    \label{fig:lateralmoment}
\end{figure}

To highlight the importance of modeling the yaw moment created from the hierarchical approach to lateral brake proportioning, comparative tests with and without the $\tau_{bb}$ term in the vehicle model were conducted on the track. In these tests, the vehicle is run through turn 3 of the Thunderhill West track. This turn has heavy simultaneous braking and steering, as is often seen during trail braking and emergency maneuvers; the results are shown in Fig. \ref{fig:lateralmoment}.

The top left plot of Fig. \ref{fig:lateralmoment} shows the estimated yaw moment (black) compared to the modeled yaw moment $\tau_{bb}$ (blue). The top right plot is from the comparison test where $\tau_{bb}$ is not modeled and set to 0 (red). When $\tau_{bb}$ is included in the model, the vehicle is able to complete the turn with much higher performance in several key metrics. The error trace in the middle left plot shows the vehicle without the yaw moment model (red) exceeds track bounds at $s = 105m$, indicating that the unmodeled moment is preventing the car from rotating in the direction of the turn, and the controller does not have the means to correctly compensate quickly enough. This is made more apparent in the yaw rate and sideslip plots in the bottom right. The controller with the yaw moment model (blue) builds sideslip and increases yaw rate much earlier in the turn. Focusing on the sideslip plot at $s = 25m$, the vehicle with yaw moment modeled builds sideslip about $5m$ before the vehicle without yaw moment modeled, which enables the car to stay within the track limits. Finally, the velocity plot shows that the vehicle is able to maintain higher velocity through both the corner entry and exit.

\subsection{UKF Adaptive MPC}

\begin{figure}
    \centering
    \includegraphics[scale = 0.75]{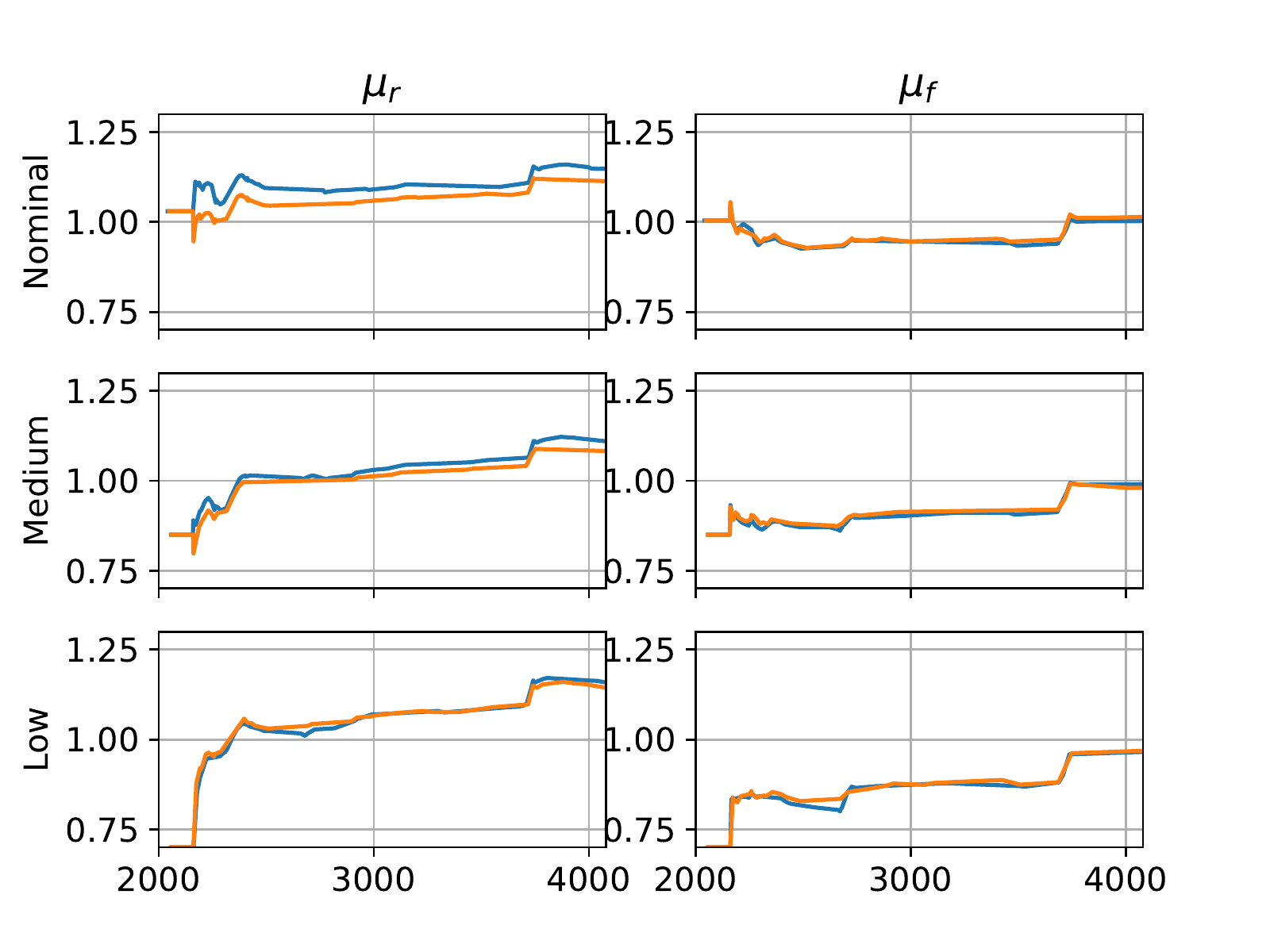}
    \caption{Experiments with in-the-loop UKF tire friction estimation with nominal (top), medium (middle), and low (bottom) initializations on Thunderhill West demonstrating accuracy and repeatability of the UKF. Orange and blue lines within each plot show separate runs with the same initial condition.}
    \label{fig:fric_nom}
\end{figure}

While allocating brakes based upon load transfer, and accounting for the additional yaw moment it creates improves the performance and operating range of NMPC, utilizing all the available friction force requires knowledge of friction and how friction varies throughout operation. Addressing this, three separate experiments, each consisting of two laps of the track under autonomous control, were conducted with different initializations of the UKF estimator: low $(\mu_f, \mu_r) = (0.7, 0.7)$, medium $(\mu_f, \mu_r) = (0.85, 0.85)$, and nominal $(\mu_f, \mu_r) = (1.003, 1.04)$. In each of these cases, the estimated friction is used to update the NMPC controller in real time. This shows the efficacy of this approach in closed-loop.

\subsubsection{Stability of UKF Estimation}
Fig. \ref{fig:fric_nom} shows the estimated friction vs. path distance for the six separate UKF runs where NMPC is updated in real-time with the estimated friction coefficients.  The top row depicts nominal initialization of $(\mu_f, \mu_r) = (1.003, 1.04)$, the middle row depicts a medium initialization of $(\mu_f, \mu_r) = (0.85, 0.85)$, while the bottom is a low initialization of $(\mu_f, \mu_r) = (0.7, 0.7)$.  Each row shows two separate runs (orange and blue) for the rear (left) and front friction (right). In all six runs, the estimator shows repeatability and captures similar trends, such as front tire friction being less than the rear. Furthermore, the UKF converges near similar values despite the low initialization yielding less lateral dynamic excitation. Slight differences are observed between individual runs which could be attributable to increased tire temperature due to ambient temperature changes throughout the day or temperature changes from repeated testing. 

\subsubsection{Adaptive NMPC: Conservative Scenario}
\begin{figure}
    \centering
    \includegraphics[scale = 0.75]{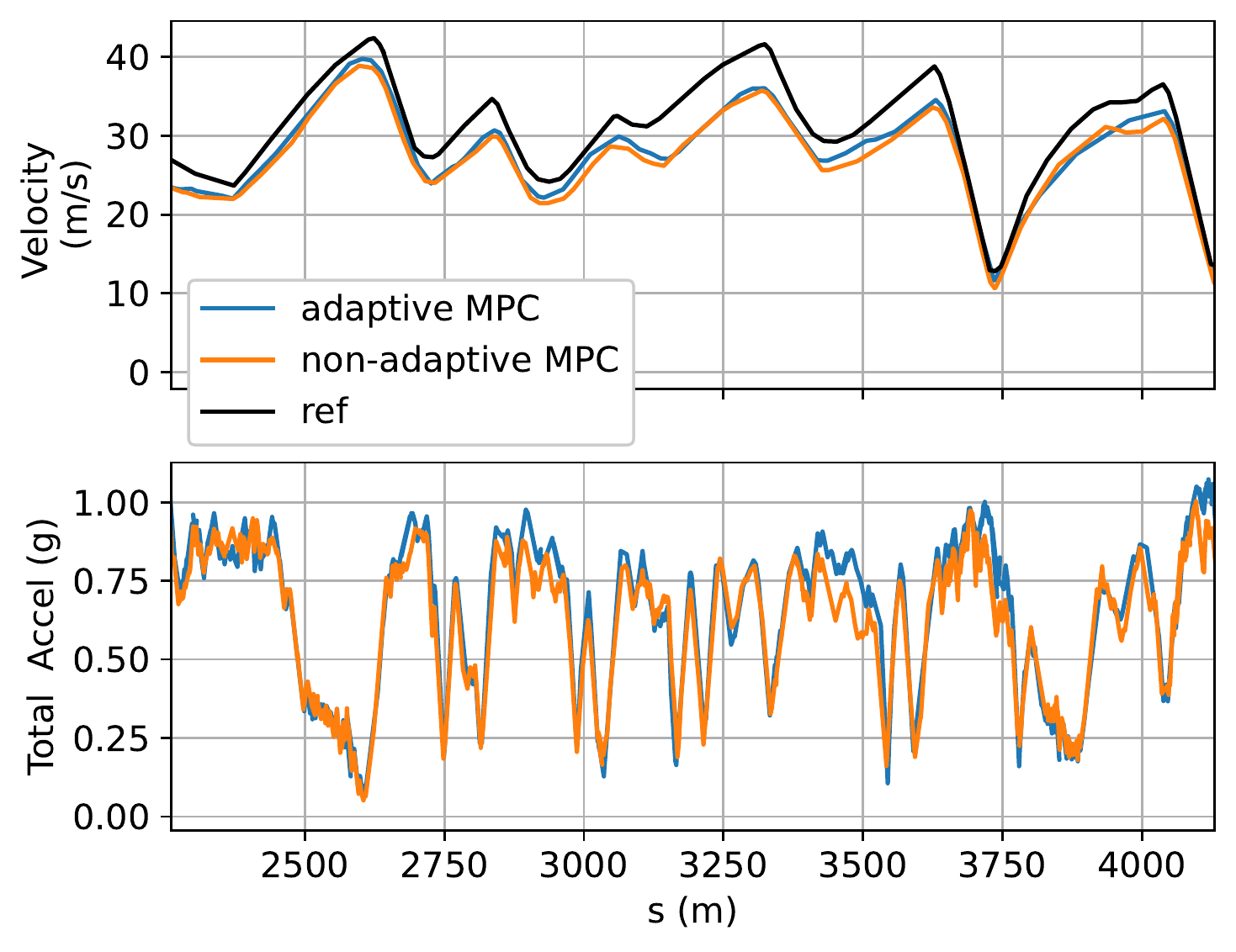}
    \caption{Thunderhill 2-mile Velocity trace (top) and total g's (bottom) for adaptive MPC (blue) and non-adaptive MPC (orange).  Reference velocity is shown in black.}
    \label{fig:AMPC_mu85}
\end{figure}

Fig. \ref{fig:AMPC_mu85} depicts the performance improvement achieved by adaptive MPC as compared to a non-adaptive baseline MPC. In this test, the UKF is initialized to: $\mu_f = 0.85$ and $\mu_r = 0.85$.  For the non-adaptive case, friction is fixed at these values throughout the whole run; for the adaptive MPC case, the UKF is initialized with the same values but runs online and updates the MPC vehicle model with the estimates. For both cases, the NMPC is configured to use up to $90\%$ of this modeled friction on the track (i.e. by setting $\mu_{lim} = 0.9$).  Despite the large initialization error, the UKF converges near the optimal values (Fig. \ref{fig:fric_nom}, middle). Fig. \ref{fig:AMPC_mu85} shows the velocity trace (top) and total acceleration (bottom) for adaptive MPC (blue) and non-adaptive MPC (orange). Adaptive MPC (blue) is able to outperform the non-adaptive MPC (orange) and achieves higher speeds (top) and higher total acceleration (bottom).  In fact the mean speed for adaptive MPC is 26.9 $m/s$ whereas the mean non-adaptive speed is 26.1 $m/s$, demonstrating the improved performance achieved through adaptation.

\subsubsection{Adaptive NMPC: Nominal Scenario}
\begin{figure}
    \centering
    \includegraphics[scale = 0.75]{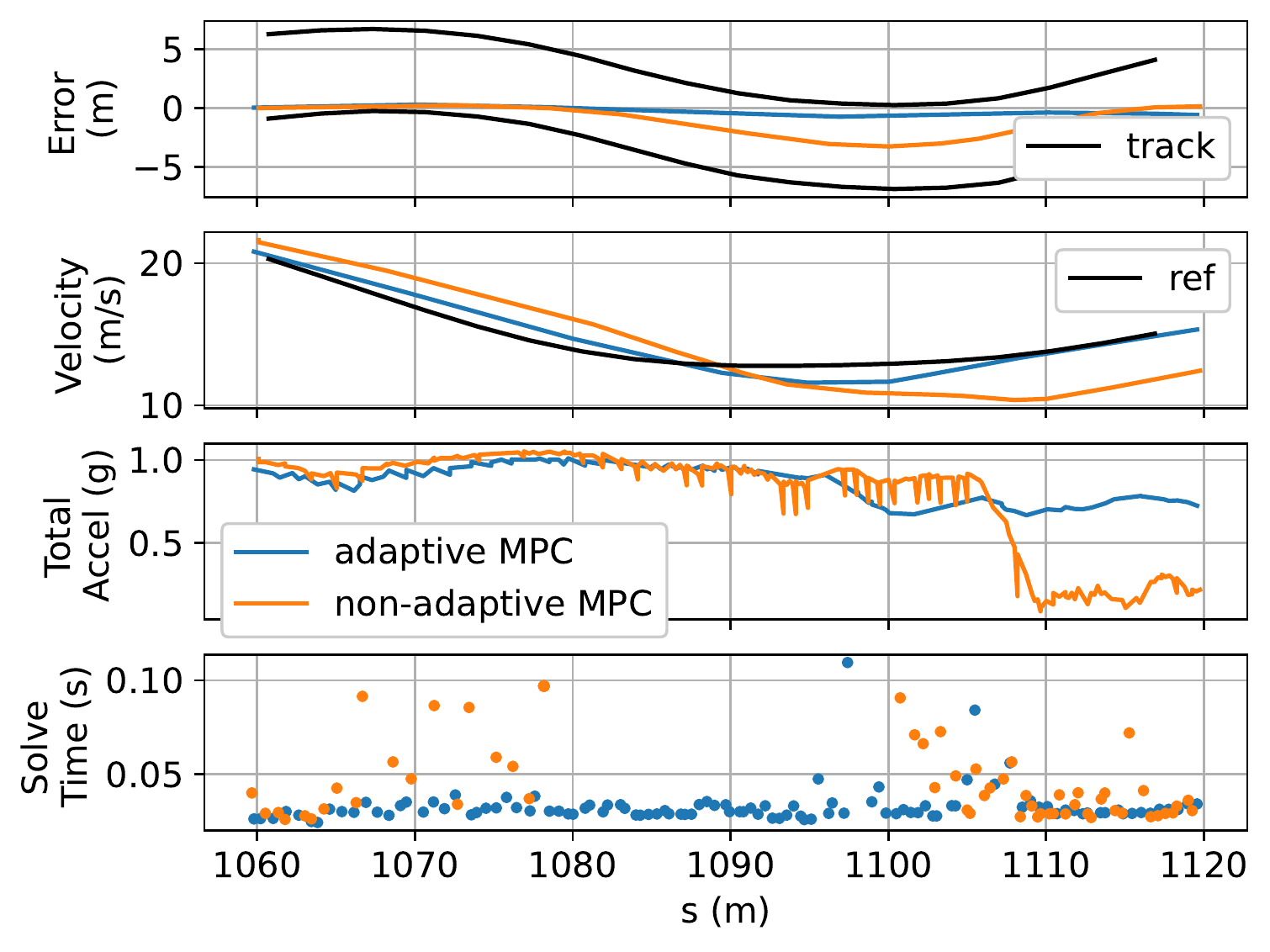}
    \caption{Autonomous tests for turn 7 of Thunderhill West.  Lateral error (top), velocity (middle), total acceleration (middle), and solve times (bottom) showing increased performance of adaptive MPC (blue) as compared to non-adaptive MPC (orange). }
    \label{fig:AMPC}
\end{figure}

Online friction estimation can also improve performance in non-conservative scenarios. In this next set of experiments, the baseline MPC is configured to run at fixed values $\mu_f = 1.004$, and $\mu_r = 1.03$, which were empirically determined to result in excellent performance across the track. This is compared to the adaptive approach which is initialized at these same friction values, but allowed to update its values online. While this improved performance throughout the track, the difference is particularly pronounced at sharp hairpin corners. In Fig. \ref{fig:AMPC}, we show the result for Turn 7 at Thunderhill West. Turn 7 is a sharp left hairpin corner where the vehicle approaches at a speed of approximately $36 m/s$ before braking to about $13 m/s$.  This also corresponds to the jump in estimated friction seen at $s \approx 3740m$ in Fig. \ref{fig:fric_nom}. The impact of online friction adaptation is visible in several aspects of this plot.  First, without friction estimation and adaptation, the non-adaptive MPC (orange) formulation operates at a peak total acceleration of 1.04g due to a overestimation of the available friction and carries more speed into the turn, as compared to 1.0g for adaptive MPC (blue) where the entry speed is more appropriate (third plot). This leads to tire saturation with non-adaptive MPC causing the vehicle to miss the apex shown by the large lateral error of the non-adaptive case (orange) of the top plot. 
The non-adaptive case (orange) also reduces speed later in the turn, as indicated by the decreased speed between s values of 1090 m and 1110 m of the second plot. Due to this saturation, non-adaptive MPC fails to converge as there is no available lateral force to correct for the error and complete the turn, shown by the large gap of solves between an s value of 1080 and 1100 m in Fig. \ref{fig:AMPC} (bottom). Adaptive MPC is able to complete the turn at the limit of traction but without saturating the tire. Furthermore, the adaptive MPC achieves an average and peak absolute lateral error of 0.43 and 0.77 $m$ which is less than the 1.4 and 3.26 $m$ of non-adaptive MPC,  Fig. \ref{fig:AMPC} (top). In both implementations, lateral error bound violations are extremely rare and adaptive MPC led to less track bound violation (0.8\%) as compared to the non-adaptive MPC (1.8\%) for the whole tested track. 

The improved performance allows adaptive MPC to carry a better speed profile through the turn allowing for earlier throttle application leading to an exit speed of 15.37 $m/s$ as compared to 12.5 $m/s$ for non-adaptive MPC. This is particularly prominent for s values after 1090 m of the second plot. In contrast to non-adaptive MPC, the ability of adaptive MPC to learn and adapt to friction allows for successful completion of turn 7 while improving robustness and performance. Complete lap time cannot be calculated because a section of the test circuit is unsuitable for autonomous operation; however, calculating the partial lap time between turn 2 and turn 9 results in 79.1s for non-adaptive MPC and 77.6s for adaptive MPC, further demonstrating improved performance.

\section{Conclusion}
Developing autonomous vehicles capable of operating at, or beyond, the limits of handling requires models that can capture complex nonlinear dynamics and accurate knowledge of environment variables. To addresses this need, this work presents a novel NMPC formulation that brings selected important chassis control functionality into the higher level vehicle model. Specifically, the predictive control layer uses a single-track model with longitudinal weight transfer dynamics and independent allocation of front and rear axle brake torques, but delegates lateral brake balance to the chassis layer. This expands the vehicle operating range by optimally allocating the brake distribution to account for the additional or reduced load at each wheel, whilst still allowing for sufficient horizon length in the predictive control layer. Next, to address the need for accurate knowledge of environment variables, UKF friction estimation is used to update the vehicle model within NMPC in real-time. This is shown to significantly increase closed loop NMPC performance. Experimental validation is performed at a friction limit of 0.95$\cdot \mu$ on a closed course track demonstrating the effectiveness of the dynamic brake balance and online friction adaptation for extracting maximum performance of the autonomous vehicle. 

This demonstrates a step towards realizing autonomous vehicles capable of utilizing the vehicle’s full capabilities when the need arises. Future work in this vein could explore further improvements to the formulation, such as assimilating more chassis level control into NMPC, incorporating friction estimate uncertainty, and the impact of estimating other parameters simultaneously. Furthermore, while the approach has been validated extensively through experiments, formal stability and recursive feasibility analysis is an important future research direction.

\bibliographystyle{apalike}
\bibliography{root}

\end{document}